\documentclass[twocolumn]{aastex701}
\usepackage{xcolor}
\usepackage{soul}
\usepackage{subcaption}
\usepackage{natbib}

\begin{document}

\title{A Taxonomic Characterization of Targets in the Emirates Mission to the Asteroid Belt}

\author[orcid=0009-0003-9667-9483,sname='Lee']{Laura M. Lee}
\affiliation{Department of Astronomy and Planetary Science, Northern Arizona University, S San Francisco St., Flagstaff, AZ 86011, USA}
\affiliation{Radiant Center for Remote Sensing, Northern Arizona University, S San Francisco St., Flagstaff, AZ 86011, USA}
\email[show]{lml446@nau.edu}  

\author[orcid=0000-0003-4580-3790,sname='Trilling']{David E. Trilling} 
\affiliation{Department of Astronomy and Planetary Science, Northern Arizona University, S San Francisco St., Flagstaff, AZ 86011, USA}
\email{david.trilling@nau.edu}

\author[orcid=0000-0003-1008-7499,sname='Kareta']{Theodore Kareta}
\affiliation{Lowell Observatory, 1400 W Mars Hill Rd., Flagstaff, AZ 86001, USA}
\affiliation{Villanova University, 800 Lancaster Ave., Villanova, PA 19085, USA}
\email{theodore.kareta@villanova.edu}

\author[orcid=0000-0003-3091-5757,sname=Thomas]{Cristina Thomas}
\affiliation{Department of Astronomy and Planetary Science, Northern Arizona University, S San Francisco St., Flagstaff, AZ 86011, USA}
\email{cristina.thomas@nau.edu}

\author[orcid=0000-0002-8552-158X,sname='van Belle']{Gerard van Belle}
\affiliation{Department of Astronomy and Planetary Science, Northern Arizona University, S San Francisco St., Flagstaff, AZ 86011, USA}
\email{Gerard.van-Belle@nau.edu}

\author[orcid=0000-0002-8096-9633,sname='Edwards']{Christopher S. Edwards}
\affiliation{Department of Astronomy and Planetary Science, Northern Arizona University, S San Francisco St., Flagstaff, AZ 86011, USA}
\affiliation{Radiant Center for Remote Sensing, Northern Arizona University, S San Francisco St., Flagstaff, AZ 86011, USA}
\email{christopher.edwards@nau.edu}

\author[orcid=0000-0003-1174-7313,sname='Haberle']{Christopher Haberle}
\affiliation{Radiant Center for Remote Sensing, Northern Arizona University, S San Francisco St., Flagstaff, AZ 86011, USA}
\email{christopher.haberle@nau.edu}

%\collaboration{all}{The Terra Mater collaboration}

\begin{abstract}
The Emirates Mission to the Asteroid Belt (EMA) will investigate seven main-belt objects to improve our understanding of the distribution and evolution of water and metal-rich materials within the asteroid population. Understanding asteroid properties prior to the mission helps anticipate potential surface activity or hazards and optimize data acquisition during brief flyby windows, while also providing essential context for interpreting in-situ measurements within the broader asteroid population. We present visible-wavelength spectra (0.40--0.95~$\mu$m) and corresponding taxonomic classifications for all seven EMA targets \added{using $\chi^2$ comparison with the visible portion of the Bus-DeMeo class templates}. Notably, our analysis identifies two B-type asteroids, a relatively rare class, providing opportunities for future compositional comparisons between EMA targets and OSIRIS-REx observations of the organic-rich asteroid Bennu. \added{We identify the remaining targets to be S-/L-, X-/C-, and D-types.} The suite of asteroids selected for EMA offers a unique opportunity to investigate both intra-class variability and compositional differences across multiple taxonomic types, advancing our understanding of chemical and mineralogical heterogeneity in the asteroid belt.

\end{abstract}



\section{Introduction} 

Asteroid research increases our understanding of the early conditions of the Solar System, including collisional histories, compositional variation, and the transport of water and the building blocks of life \added{\citep{BottkeAstIV2015, DemeoAstIV2015, ReddyAstIV2015, KrotAstIV2015}}. These \added{building blocks} formed long before the formation of the planets and most were accreted to become planetesimals and planets. Unlike planets, asteroids today are relatively unchanged from $\sim$4 billion years ago, providing clues to the conditions and processes that existed during the solar system’s formation \added{\citep{DemeoAstIV2015}}. By studying the composition of asteroids we can learn about the distribution of materials (i.e. metals, silicates, water, organics) in the early solar nebula, which influenced planet formation \added{\citep{DemeoAstIV2015, KrotAstIV2015}}. Additionally, some carbon-rich asteroids contain organic compounds and water bearing minerals. For example, analyzed samples from asteroid Ryugu (C-type) detected amino acid molecules \citep{Oba2023}, and samples from asteroid Bennu (B-type) were found to be abundant in not only carbon, nitrogen, and ammonia, but also amino acids and all nucleobases needed in the creation of RNA and DNA \citep{Glavin2025}. \added{Asteroid missions are contributing details that build on conditions for} solar system evolution, how the building blocks of life arrived on Earth, and what bodies are responsible for the presence of water on Earth \citep{morbidelli2000, alexander2018}.

\added{Remote-sensing reflectance spectroscopy is the most accessible while still highly informative way to study a large number of asteroids.} The reflectance slope and absorption features in an asteroid’s visible (V) and near-infrared (NIR) spectrum are used to classify these small bodies, providing inferences about surface composition and thermal alteration history \added{\citep{Tholen1984, BusBinzel2002,DEMEO2009, mahlke2022}}. The three most common Bus-DeMeo asteroid taxonomies are S-, X-, and C-complex. S-complex asteroids are identified \added{by} a red sloped spectrum \added{with a} 1 $\mu$m absorption feature \added{and is} thought to be stony and silicate-rich with a surface composition of pyroxene and/or olivine \added{\citep{gaffey1993, BusBinzel2002}}. X-complex asteroids are a loosely defined class that is identified by featureless spectra and a range of spectral slopes from red to relatively neutral. \added{Asteroids in the} X-complex are sometimes associated with a metallic composition and are grouped into classes \added{E, M, or P} based on the asteroid's albedo and spectral features in the infrared \added{\citep{Tholen1984,mahlke2022}}. \added{The E-class is linked to enstatite achondrites, the M-class is associated with silicate-rich metallic types, and P-types are carbonaceous. (\citealt{clark2004,FORNASIER2011} and references therein).} C-complex asteroids have featureless neutral-red-to-blue-sloped visible spectra and are believed to be primitive objects rich in carbon, hydrated minerals, and possibly organics \citep{gaffey1978, BusBinzel2002,Tholen1984, rivkin2003, rivkin2012}. A class within the C-complex are B-type asteroids. Characterized by their blue-sloped visible reflectance spectra, B-types make up $\sim$5\% of discovered asteroids and are thought to contain magnetite, graphite, phyllosilicates, and ferrihydrite \citep{yang2010,de2012,demeo2013taxonomic,Chavan2025}. \added{In some cases, targets in the X- and C-complex display an absorption feature around 0.7 $\mu$m associated to the presence of Fe-rich phyllosilicates which has been correlated to hydrated materials \citep{gaffey1978, vilas1994,rivkin2012,fornasier2014}.}

Recent asteroid missions like Hayabusa2, OSIRIS-REx, and Lucy have expanded our understanding of asteroid diversity, revealing critical details about surface homogeneity, crater morphology, and fragmentation processes \added{\citep{yada2022,lauretta2017,levison2021, barucci2020, kitazato2019}}. These findings have provided insight into the collisional history of the solar system and its role in planetary evolution \citep{Michel2009,Bierhaus2022,March2023}. By studying asteroids across different taxonomic classes, we piece together a more complete picture of the processes that have shaped our planetary system over billions of years.

The \added{United Arab Emirates Space Agency's} Emirates Mission to the Asteroid Belt (EMA) will visit 7 asteroids throughout the course of the mission \added{and will launch in 2028 \citep{Hoor2023}}. Onboard EMA is a visible (420--700 nm) and InfraRed (IR; 8--14 $\mu m$) imager (EMACS), the Emirates Main Belt IR Spectrometer (\added{EMBIRS; }6--40+ $\mu m$; \citealt{edwards2025}), the Mid-wave infrared Imaging Spectrometer for Target Asteroids (MIST-A; 2--5 $\mu m$; \citealt{filacchione2025}), an electron and proton telescope (REPTile-3), and a deployable lander equipped with a wide- and narrow-angle camera \citep{Alshemeili2025, Reed2023}. This suite of instruments will enable analysis on surface compositions (i.e. hydrated minerals, volatiles, organics), thermal properties, general geology, and surface alterations from solar energetic particles.

EMA will complete 6 asteroid flybys and end with a rendezvous and lander contact with the seventh and final target. The mission timeline includes flybys of (10253) Westerwald, (623) Chimaera, (13294) Rockox, (88055) Ghaf, (23871) Ousha, and (59980) Moza, spanning 2030 to 2033. In 2034, EMA will rendezvous with asteroid (269) Justitia, which is roughly 50 km in diameter, is one of the reddest known objects in the asteroid belt, and is thought to have originated as a transneptunian object (TNO) \citep{Hasegawa2021, Rivikin2023, Reed2023, hayne2025}. Based on \added{visible} spectroscopic similarities with other very red TNOs (RR/IR) \added{and classifications to the Z-class by \citet{mahlke2022}}, Justitia is thought to have complex organic materials on its surface \citep{Hasegawa2021}. \added{Additional observations covering the 3~$\mu$m region support the potential presence of organic materials, phyllosilicates, and hydrated minerals, however this region shows more similarities with B-type asteroids, Bennu and Polana, than TNOs \citep{Rivkin2026}.} To date, only two \added{possibly} TNO-like bodies, Justitia and (203) Pompeja, have been identified in the asteroid belt \citep{Hasegawa2021}, making this mission a rare opportunity to not only study multiple asteroids, but also to \added{potentially} gain deeper insight into outer solar system material that has migrated inward. \added{The mission goal of this main belt asteroid tour is} to further our understanding on where and how hydrous asteroids formed and how these bodies have evolved over time. Table~\ref{tab:param_table} shows a compilation of the publicly available information on these targets prior to this study. Taxonomic information is known for some of these targets, though not all, and not all published results agree. At the time of writing, asteroid (13294) Rockox had yet to be classified \added{via spectroscopy}, but has an albedo of 0.14 \citep{masiero2011} and Sloan Digital Sky Survey (SDSS) colors that point to a CX- or S-type asteroid \citep{carvano2010,demeo2013taxonomic}. Asteroid (88055) Ghaf had no previously identified taxonomic type.

%\centering
\begin{table*}[htb!]
\centering
%\label{tab:param_table}
\begin{tabular}{llllllll}
\hline
Asteroid&H mag [1]&Diameter&Albedo&a (AU) [4]&Rotation&Lightcurve&Taxonomic Type\\
 &&(km)& && Period (hrs) & Amp (mag) &\\
\hline
269 \added{Justitia} & 9.8 & 50.7 [2]& 0.06 $\pm$ 0.01 [2]& 2.613 & 33 [5,6]& 0.19 [5] & D, Ld, \added{Z}\\
623 \added{Chimaera} & 10.8 & 44.1 [3]& 0.037 $\pm$ 0.002 [3]& 2.461 & 14.6 [6]& 0.2 [6]& \added{Ch, Xc,} X or C\\
10253 \added{Westerwald} & 15.3 & 2.27 [2]& 0.31 $\pm$ 0.03 [2]& 2.296 & 12.9 [5]& 0.12 [5]& S\\
13294 \added{Rockox} & 14.4 & 5.2 [2]& 0.14 $\pm$ 0.02 [2]& 2.257 & 3.5 [5]& 0.2 [5]& S, CX, \added{X}\\
23871 \added{Ousha} & 15 & 6.7 [2]& 0.05 $\pm$ 0.03 [2]& 2.387 & --- & --- & C\\
59980 \added{Moza} & 13.3 & 8 [2]& 0.12 $\pm$ 0.02 [2]& 3.054 & (5.3) [5] & (0.3) [5] & X, C, L, \added{S}\\
88055 \added{Ghaf} & 15.3 & 5.4 [2]& 0.05 $\pm$ 0.01 [2]& 2.734 & (2.6) [5] & (0.1) [5] & --- \\
%Consider using \footnote{\url{http://www.latex-project.org/}} for sources here

\hline
\end{tabular}
\caption{Previously reported parameters of each target from various studies order\added{ed by ascending asteroid number. References for taxonomic type are Justitia [7-9,10,11], Chimaera [11,12,13,13], Westerwald [14-16], Rockox [14,15,17], Ousha [15-17], and Moza [14,15,15,17]}. [1] Minor Planet Center \citep{MPC}, [2] NEOWISE \citep{masiero2011, masiero2014, masiero2017}, [3] Infrared Astronomical Satellite \citep{IRAS}, [4] NASA JPL Solar System Dynamics - Small-Body Database, [5] SNAPS \citep{Trilling2023}, [6] Asteroid Lightcurve Database \citep{LCDB2009}, [7] \citet{DEMEO2009}, [8] \citet{Hasegawa2021}, [9] \citet{Bus1999}, [10] Small Main-Belt Asteroid Spectroscopic Survey, Phase II \citep{SMASSII2020}, [11] \citet{mahlke2022}, \added{[12]\citet{Morate2019}, }[13] \citet{tholen1989}, [14] \citet{demeo2013taxonomic}, [15] \citet{carvano2010},\added{ [16]\citet{sergeyev2021}, [17]\citet{sergeyev2022}}.
Values in parentheses indicate solutions that are marginal or low significance.
}
\label{tab:param_table}
\end{table*}

Multi-body missions have inherent limitations, and basic characterization data may be lacking for some targets. For EMA, target selection was driven by optimal trajectory to visit seven asteroids, with Justitia as the primary target. Consequently, several targets remain understudied, particularly the smaller bodies (Table~\ref{tab:param_table}). Here we present a comprehensive \added{visible spectroscopic} study of all seven EMA targets in order of mission encounter \added{using the DeVeny instrument at the Lowell Discovery Telescope}. In Section~\ref{sec:obs_data} we present our observations and data processing. Section~\ref{sec:results} presents our results using the Bus taxonomy classification system \added{\citep{BusBinzel2002}}. Section~\ref{sec:disc} discusses our findings and inferences on surface composition. In Section~\ref{sec:conclusion}, we present our conclusions and future work.

\section{Observations and Data Processing} \label{sec:obs_data}
\subsection{Observations}
We collected 0.40--0.95 $\mu m$ spectra of the EMA targets using the DeVeny instrument \citep{Ellsworth2023} on the 4.3-meter Lowell Discovery Telescope (LDT) between 2023 December 4 -- 2025 August 20 UT. Observations were made using a 3.0\arcsec ~slit with the DV1 grating, and no order blocking filter. \added{Though second-order light may have minor effects on data longward of 0.85 $\mu$m, with DeVeny, the overall spectral slope and the presence of a 1 $\mu$m feature is not largely affected}. Under these conditions, the spectral resolution (R) is 450 and the groove density is 150~g~$mm^{-1}$. At the beginning of each observing run, standard calibration procedures for DeVeny were followed. During science observations we regularly switched between observing the target asteroid and a nearby G2V standard star. Details on each object’s observing run are given in Table~\ref{tab:obs_table}. Observations were driven by when each target was observable and how much time was allotted on the telescope (i.e. half night, whole night). In most cases, targets were only visible for parts of the night, so we prioritized collecting many long exposure frames (300-500 seconds) over meeting a target signal to noise ratio value.

\begin{table*}[htb!]
\centering
%\label{tab:obs_table}
\begin{tabular}{lcccclc}
\hline
Target & Obs Date $\&$ \added{Time} [UT] & V mag & \added{Exposure} & \added{$\alpha$} & Standard Star & Airmass \\
\hline
(269) \added{Justitia} & 2023-12-04 \added{02:50}& 14.2 & \added{35x300s} & \added{6.9} & Cl Melotte 22 761 & 1.02 -- 1.54 \\
(623) \added{Chimaera} & 2025-08-20 \added{10:36}& 15.0 & \added{7x300s} & \added{24.3} & HD 11811 & 1.00 -- 1.02 \\
(10253) \added{Westerwald} & 2025-08-20 \added{09:31}& 18.8 & \added{1x300s,
6x500s} & \added{8.3} & HD 220764 $\&$ & 1.32 -- 1.95 \\
 & & & & &BD07595988&\\
(13294) \added{Rockox} & 2024-02-15 \added{06:29}& 17.0 & \added{11x300s} & \added{10.9} & HD 96868 & 1.12 -- 1.39 \\
(23871) \added{Ousha} & 2024-11-09 \added{03:27}& 18.4 & \added{1x300s,  24x400s} & \added{5.8} & GAT 1048 & 1.14 -- 2.04 \\
(59980) \added{Moza} & 2024-04-09 \added{05:51}& 18.1 & \added{15x300s} & \added{9.9} & HD 133623 & 1.20 -- 2.26 \\
(88055) \added{Ghaf} & 2023-12-04 \added{06:48}& 19.6 & \added{49x300s} & \added{14.4} & HD 257880 & 1.00 -- 1.51 \\
%Consider using \footnote{\url{http://www.latex-project.org/}} for sources here
\hline
\end{tabular}
\caption{Summary of our observations of the EMA targets \added{in ascending} order \added{by target number}. \added{Time shown represent target observing start time.} V magnitudes \added{and phase angle ($\alpha$)} are reported from NASA JPL Solar System Dynamics - Horizons System.}
\label{tab:obs_table}
\end{table*}

In this paper we also report \added{Sloan g', r', i', z'} spectrophotometric measurements of (269) Justitia that were obtained using the 0.5-meter Titan Monitor telescope (TiMo; \citealt{van2020}) at Lowell Observatory’s campus on Mars Hill in Flagstaff, Arizona, during 2025 February 27-28 and March 1 and 5 UT. \added{Observations were collected using 300 second exposures in a g-r-i-z observing sequence.}

\subsection{Data Reduction}
We used the Python spectroscopy package PypeIt \citep{Prochaska2020} to reduce our spectroscopic data. PypeIt reads in spectroscopic data, distinguishing between arcs, biases, flats, and science images, and corrects and calibrates the science images using the provided arcs, biases, and flats. Once reduced, the 1D spectrum of each science exposure is produced.

After flatfielding, debiasing, and wavelength calibrating the data in PypeIt, we then coadded the extracted spectra of each standard star taken at similar airmass conditions to produce the average spectrum of the standard star. In other words, if a standard star was observed at high and low airmasses at different points in the night, they are combined separately in blocks. Next, each spectrum of the asteroid is divided by the average spectrum of the standard star with closest airmass conditions. Our pipeline can optionally apply an extinction curve to the individual frames of the standards and asteroids before averaging and combination, but as long as they were taken at similar airmass this step does not meaningfully change the retrieved signal to noise or spectral slope \citep{Kareta2025}. This ratio spectrum of an asteroid divided by a nearby (G2V) solar analog approximately produces a reflectance spectrum of the asteroid. The asteroid’s reflectance spectra are then averaged together, normalized at 0.55 um, and used in further analysis. 

Observations collected by TiMo were calibrated and processed into photometric data using PHOTOMETRYPIPELINE\footnote{\url{https://photometrypipeline.readthedocs.io/en/latest/}} (PP; \citealt{ Mommert2017}). PP calibrates its photometry through a query of the Sloan Digital Sky Survey Data Release 9 (SDSS-R9; \citealt{ahn2012}) for background field stars that come in each image; this approach mitigates transparency and/or seeing variations between nights or facilities. Processed magnitude data were then averaged over each hour of observation, the SDSS solar contributions were removed using values from \citet{Willmer2018}, and magnitudes were converted to reflectance and normalized to the r’ band value produced in Justitia’s DeVeny reflectance spectrum.

\section{Results} \label{sec:results}

\begin{figure*}[htb!]
\centering
%\plotone{AllAsteroidDot.png}
\includegraphics[scale=0.50]{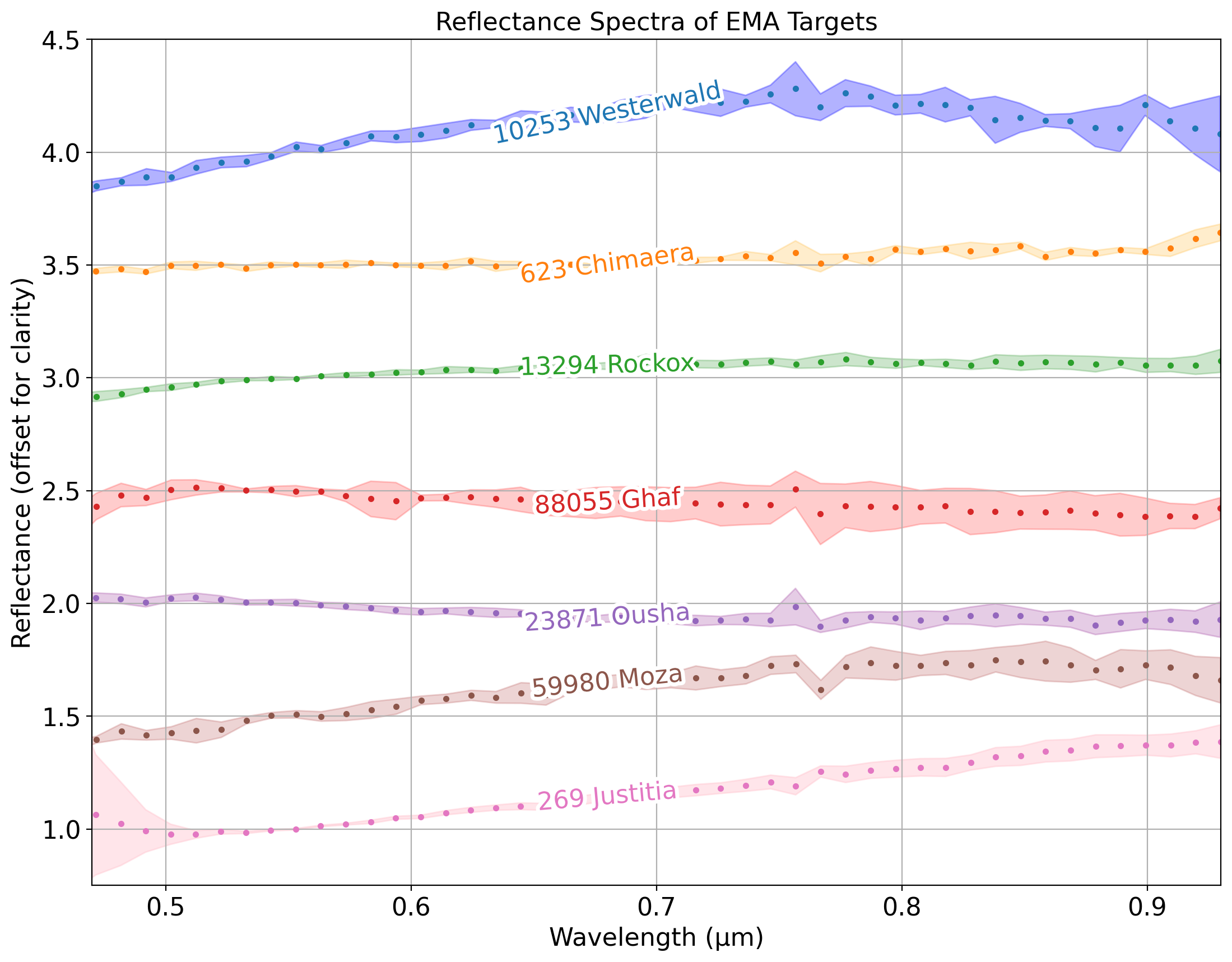}
\caption{Average reflectance spectra of all seven EMA targets ordered based on EMA visit. Reflectance spectra have been normalized at 0.55 $\mu m$ and are offset by 0.5 for Moza and Ousha and in increments of 0.25 for the remaining targets for clarity. Shading represents the standard \added{error} of each \added{observed} spectrum.}
\label{fig:allast}
\end{figure*}

\begin{figure}[htb]
\includegraphics[scale=0.6]{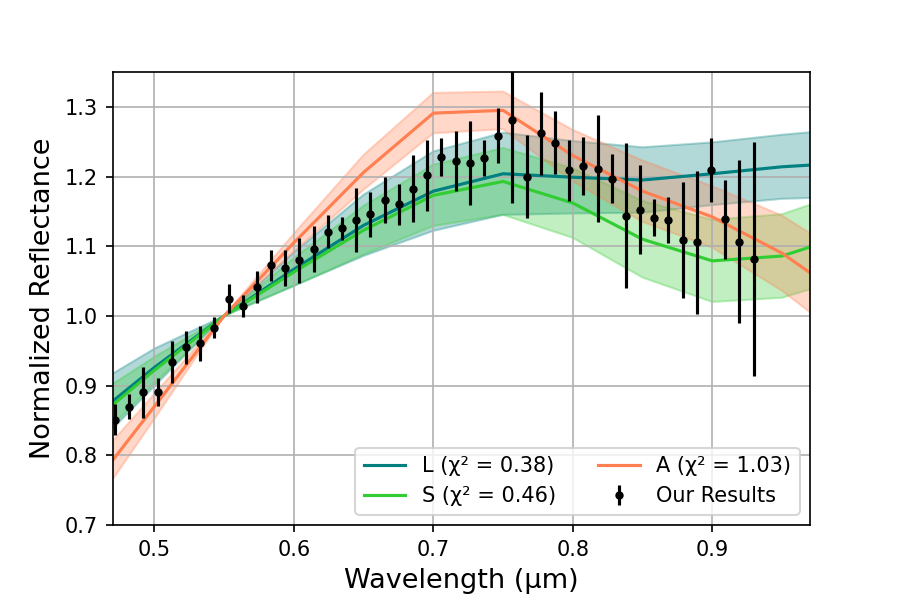}
%\plotone{10253.png}
\caption{Spectrum of Asteroid (10253) Westerwald plotted in black with bars to represent the standard \added{error} of the \added{observed} spectrum. Similar Bus taxonomies are plotted as solid colored lines with shading to represent class range. We find that this asteroid is most likely an S-type.}
\label{fig:10253}
\end{figure}

\begin{figure}[htb]
\includegraphics[scale=0.6]{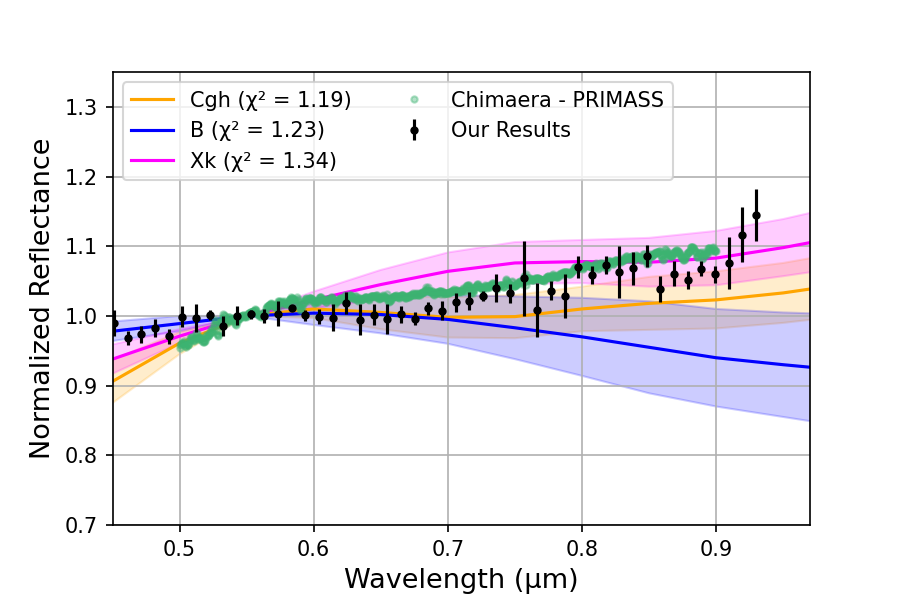}
%\plotone{623.png}
\caption{Spectrum of Asteroid (623) Chimaera plotted in black with bars to represent the standard \added{error} of the \added{observed} spectrum. Similar Bus taxonomies are plotted as solid colored lines with shading to represent class range. \added{The published spectrum of Chimaera from the PRIMASS survey is plotted in green for comparison \citep{Morate2019}.} We find that this asteroid is most likely a C- or X-type.}
\label{fig:623}
\end{figure}

\begin{figure}[htb]
\includegraphics[scale=0.6]{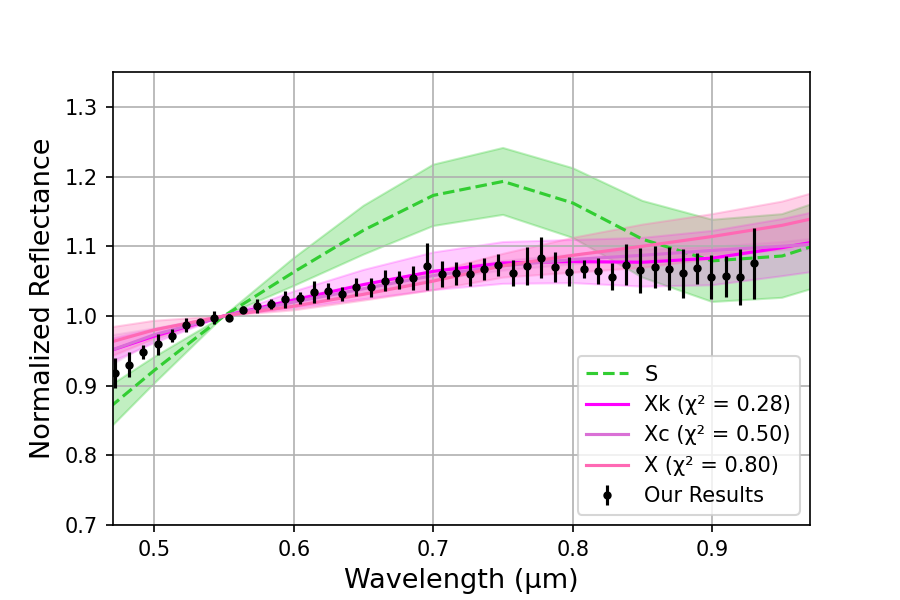}
%\plotone{13294.png}
\caption{Spectrum of Asteroid (13294) Rockox plotted in black with bars to represent the standard \added{error} of the \added{observed} spectrum. Similar Bus taxonomies are plotted as solid colored lines with shading to represent class range. For classes with no shading, no range was provided. \added{Classes} within the same complex are plotted in similar colors. The previous \added{S-type} classification \added{(via photometry)} that is not in agreement with our results is plotted as a dashed line. We find that this asteroid is most likely an X-type \added{in agreement with photometry surveys by \citet{carvano2010} and \citet{sergeyev2022}}.}
\label{fig:13294}
\end{figure}

\begin{figure}[htb]
\includegraphics[scale=0.6]{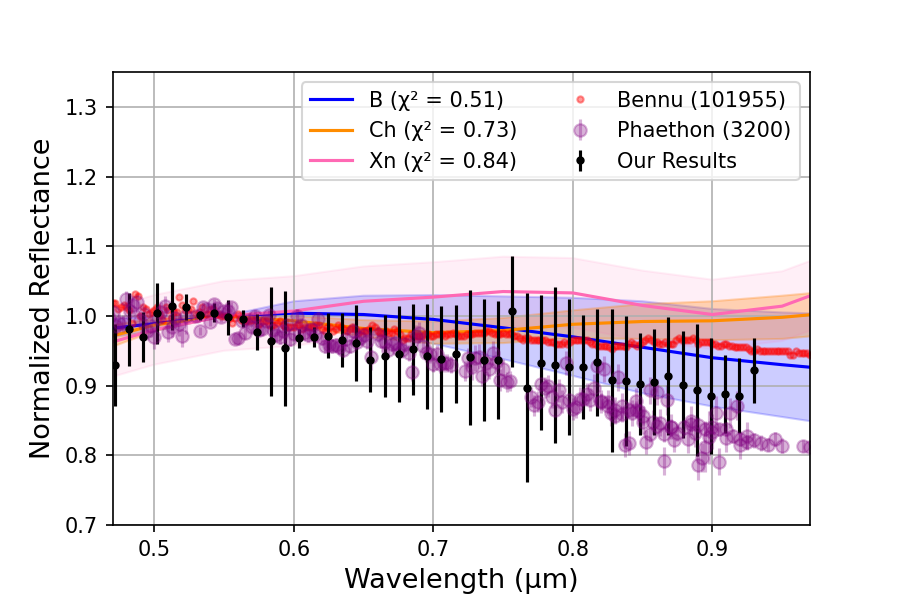}
%\plotone{88055.png}
\caption{Spectrum of Asteroid (88055) Ghaf plotted in black with bars to represent the standard \added{error} of the \added{observed} spectrum. Similar Bus taxonomies are plotted as solid colored lines with shading to represent class range. We find that this asteroid is most likely a B-type. Spectra of Bennu (red; \citealt{hamilton2019}) and Phaethon (purple; \citealt{BusBinzel2002}) included for B-type comparisons.}
\label{fig:88055}
\end{figure}

\begin{figure}[htb]
\includegraphics[scale=0.6]{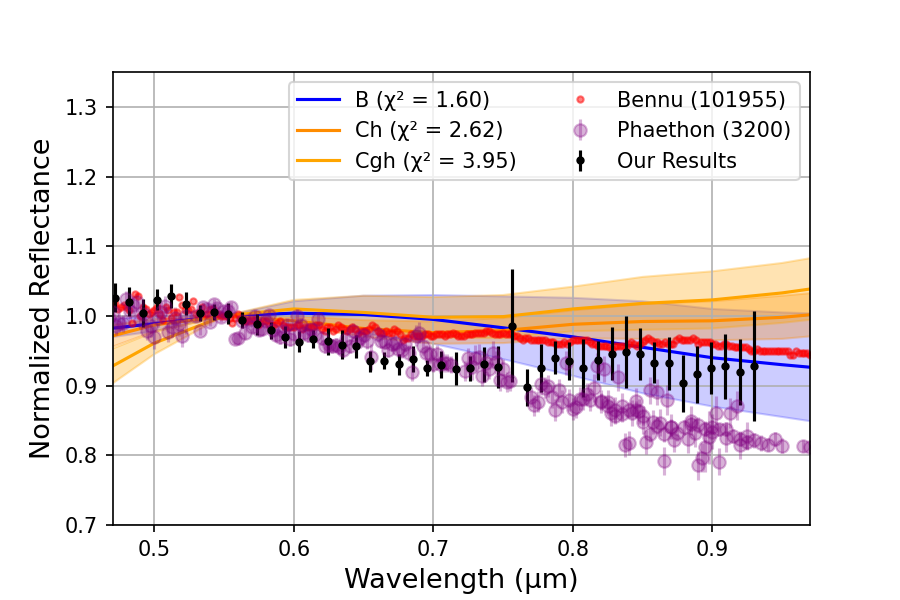}
%\plotone{23871.png}
\caption{Spectrum of Asteroid (23871) Ousha plotted in black with bars to represent the standard \added{error} of the \added{observed} spectrum. Similar Bus taxonomies are plotted as solid colored lines with shading to represent class range. We find that this asteroid is most likely a B-type. Spectra of Bennu (red; \citealt{hamilton2019}) and Phaethon (purple; \citealt{BusBinzel2002}) included for B-type comparisons.}
\label{fig:23871}
\end{figure}

\begin{figure}[htb]
\includegraphics[scale=0.6]{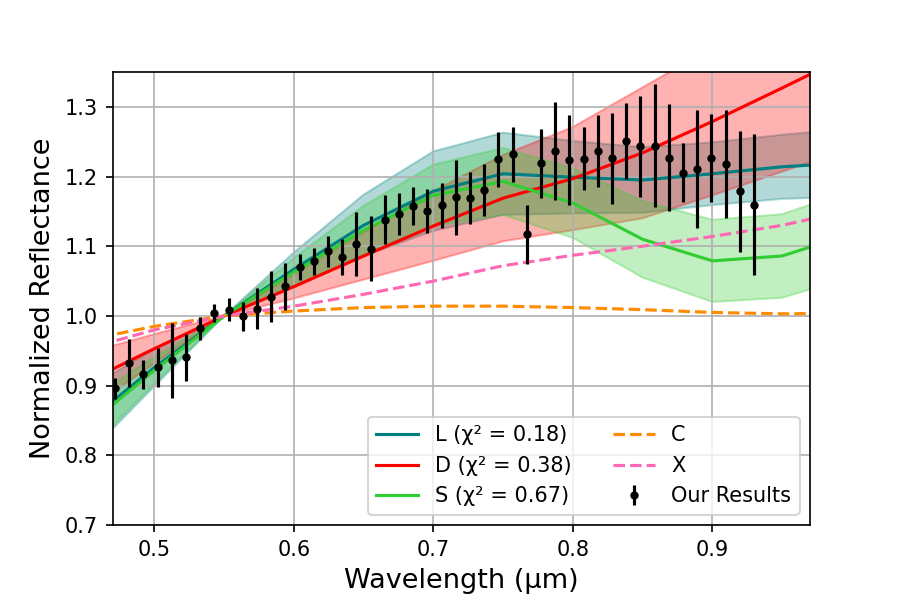}
%\plotone{59980.png}
\caption{Spectrum of Asteroid (59980) Moza plotted in black with bars to represent the standard \added{error} of the \added{observed} spectrum. Similar Bus taxonomies are plotted as solid colored lines with shading to represent class range. Previous classifications that are not in agreement with our results are plotted as dashed lines. For classes with no shading, no range was provided. We find that this asteroid is most likely an S- or L-type.}
\label{fig:59980}
\end{figure}

\begin{figure}[htb]
\includegraphics[scale=0.6]{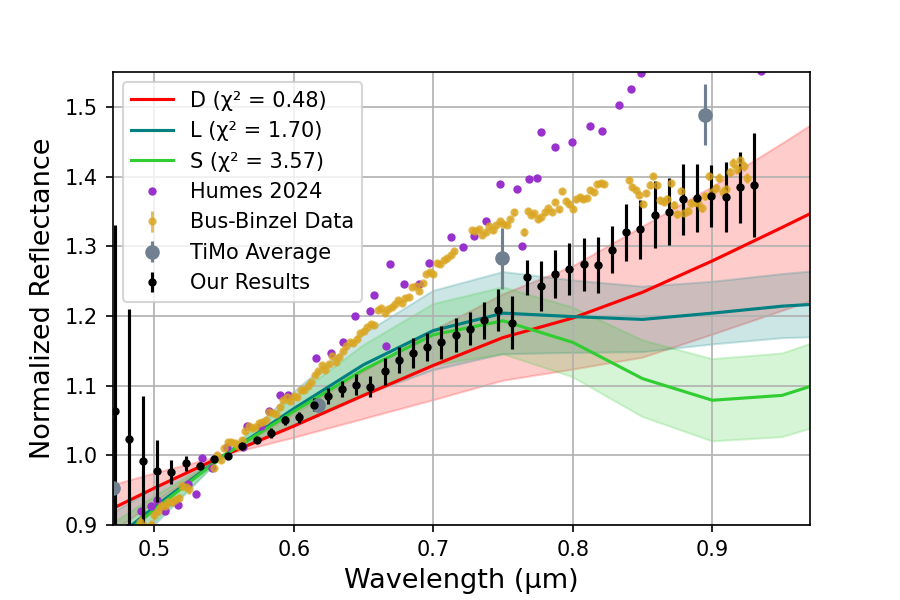}
%\plotone{269.png}
\caption{Spectrum of Asteroid (269) Justitia plotted in black with bars to represent the standard error of the observed spectrum. Reflectance-converted griz data collected by TiMo is plotted in gray with bars to represent the standard deviation of the longitudinal coverage. Similar Bus taxonomies plotted as solid colored lines with shading to represent class range. \added{The published spectrum of Justitia from \citet{humes2024} and \citet{BusBinzel2002} is plotted in purpler and gold respectively for comparison.} We find that this asteroid is most likely a D-type.}
\label{fig:269}
\end{figure}

The observed EMA objects were classified using their averaged visible spectra, applying the \added{\citet{BusBinzel2002}} taxonomic classification system. We applied the \added{\citet{BusBinzel2002}} taxonomy framework by systematically comparing each asteroid spectrum to the reference spectral classes \added{reduced} from \citet{DEMEO2009} \added{to match visible coverage from \citet{BusBinzel2002}.} The best match was assessed based on spectral slope, absorption band position, and overall continuum shape \added{through reduced $\chi^2$ goodness of fit comparisons where we include both the spectral uncertainty and DeMeo class range in the denominator}. Each asteroid's spectrum is shown in Figure~\ref{fig:allast} and their corresponding Bus classification is shown in Figures~\ref{fig:10253}--\ref{fig:269}. \added{Spectra are available in Tables~\ref{tab:WesterwaldData}--\ref{tab:JustitiaData} of the appendix. Spectral slope and absorption features at 0.7~$\mu$m were measured and identified using the CANA tool, with results presented in Table~\ref{tab:CANA} \citep{DePra2018}.} Target classification comparison can be found in Table~\ref{tab:tax_table}.

Asteroid (10253) Westerwald, an inner main belt object, was previously classified as an S-type by visible photometry using SDSS colors \added{\citep{demeo2013taxonomic, carvano2010, sergeyev2021}}. Our observed reflectance spectrum of Westerwald (Figure~\ref{fig:10253}) \added{best fits L- and S-type spectra, though the presence of the pyroxene feature at 0.9 $\mu$m is characteristic of an S-type}. Westerwald's reported albedo of 0.31 +/- 0.03 is most consistent with an S-type classification \citep{masiero2011, demeo2013taxonomic}.

Asteroid (623) Chimaera, an inner main belt object that is largest in the Chimaera asteroid family \added{\citep{Nesvorny2024, Morate2019}}, was previously classified as an X- or C-type by \citet{tholen1989} using visible photometry \added{and a Ch-type and Xc-type by \citet{mahlke2022} and \citet{Morate2019}, respectively, using spectroscopy}. Our observed reflectance spectrum of Chimaera (Figure~\ref{fig:623}) is consistent with Bus’ C- and X-complex spectra. Chimaera has an albedo of 0.04 which is consistent with albedos in the C- and X-complex \citep{IRAS, demeo2013taxonomic}. Within the X-complex, this low albedo is in agreement with P-types. \added{The spectrum of Chimaera shows a small feature around 0.7 $\mu$m and has been observed in the Chimaera family by \citet{Morate2019}. Using the CANA tool, the band depth is significant enough to meet absorption band criteria (Figure~\ref{tab:CANA}; \citealt{DePra2018}). Additionally, there is an upturn in the spectrum longward of 0.9 $\mu$m, likely caused by systematic noise of the detector.}

Asteroid (13294) Rockox, an inner main belt object in the Flora family \citep{Nesvorny2024}, is a previously unclassified object \added{via spectroscopy}, but has an albedo of 0.14 \citep{masiero2011}, SDSS colors from visible photometry that imply an S- or CX-type asteroid \citep{demeo2013taxonomic, carvano2010} \added{, and SkyMAPPER photometry that shows Rockox as an X-type \citep{sergeyev2022}}. Members in the Flora family have been classified has S-, C-, and less commonly X-types \citep{erasmus2019}. Our observed reflectance spectrum of Rockox compared to the Bus S-type classification does not show a strong 1-$\mu m$ silicate absorption feature (Figure~\ref{fig:13294}). This indicates that Rockox is unlikely to be an S-type asteroid and instead is more consistent with X-types. Rockox has an albedo of 0.14 which is consistent with M-types in the X-complex \citep{masiero2011, demeo2013taxonomic}.

Asteroid (88055) Ghaf, a middle belt object with no previous classification, shows a steep blue-sloped spectrum indicative of a B-type asteroid, part of the larger C-complex (Figure~\ref{fig:88055}). The reflectance of Ghaf tracks between that of (3200) Phaethon \citep{BusBinzel2002} and (101955) Bennu \citep{hamilton2019}, discussed further in Section~\ref{sec:disc}. This C-complex classification of Ghaf is also in agreement with its recorded albedo of 0.05 \citep{masiero2011, demeo2013taxonomic}.

Asteroid (23871) Ousha, an inner main belt object in the Erigone asteroid family \citep{Nesvorny2024}, was previously classified as a C-type using visible photometry \added{\citep{carvano2010, sergeyev2021,sergeyev2022}}. Members in the Erigone family are most commonly classified as C-, S-, B-, and X-types \citep{morate2016}. Our observations (Figure~\ref{fig:23871}) show a steep blue-sloped spectrum indicative of a B-type asteroid. Interestingly, the reflectance of Ousha also tracks between that of (3200) Phaethon from \citet{BusBinzel2002} and (101955) Bennu from \citet{hamilton2019}, discussed further in Section~\ref{sec:disc}. \added{Based on findings from the PRIMASS study, 11$\%$ of the Erigone family are B-types \citep{morate2016}. Ousha's spectrum fits within the range of B-type spectra in the Erigone family and is within Erigone's Yarkovsky cone (absolute magnitude vs. semi-major axis).} \added{The spectrum of Ousha shows a small feature around 0.7 $\mu$m and using the CANA tool, the band depth is significant enough to meet absorption band criteria (Figure~\ref{tab:CANA}; \citealt{DePra2018}). Within the Erigone family, $\sim$36$\%$ of the B-types have this absorption feature \citep{morate2016}.} This C-complex classification of Ousha is in agreement with its measured albedo of 0.05 \citep{masiero2017, demeo2013taxonomic}.

Asteroid (59980) Moza, an outer main belt object, was previously classified using SDSS colors as an X-type by \citet{demeo2013taxonomic}, a C- or L-type by \citet{carvano2010}, \added{and an S-type by \citet{sergeyev2022}, all} using visible photometry. Our observed reflectance spectrum of Moza (Figure~\ref{fig:59980}) has spectral similarities to a variety of types, notably S- and L-types, indicating that longer-wavelength observations would be required to discriminate between these types; however, we can rule out the previously hypothesized X- and C-types as likely classifications. Moza has an albedo of 0.12 which is slightly \added{darker} than general S-types, but is consistent with the L-type classification \citep{masiero2011, demeo2013taxonomic}.

Object (269) Justitia, a highly-studied middle belt object, is previously classified as a D-type by Bus-DeMeo using near-infrared spectroscopy \citep{DEMEO2009}, an Ld-type by \citet{Bus1999} and SMASSII using visible spectroscopy, and a Z-type in the \citet{mahlke2022} classification system using visible and near-infrared spectra. Our observed reflectance spectrum of Justitia (Figure~\ref{fig:269}) shows a consistent steep red-sloped spectrum indicative of a D-type asteroid \added{in the \citet{BusBinzel2002} classification system, a Z-class in the \citet{mahlke2022} classification system, }or a RR/IR-class TNO. Justitia has had albedo measurements between 0.06--0.12 \citep{masiero2014,IRAS,usui2011}, which is slightly brighter than the general D-type albedo range of 0.02--0.05 but is in greater agreement with \added{the Z-class range (0.4--0.11) by \citet{mahlke2022} and} albedos in various TNO classes \citep{Belskaya2015}.

\begin{table*}[htb!]
\centering
%\label{tab:obs_table}
\begin{tabular}{lcccc}
\hline
Target & Slope ($\%$/0.427 $\mu$m) & Band Detected & Depth ($\%$) & Center ($\mu$m) \\
\hline
(10253) Westerwald & 5.64 $\pm$ 0.15 & -- & -- & -- \\
(623) Chimaera & 2.09 $\pm$ 0.10 & Yes & 2.51 $\pm$ 0.33 & 0.69 $\pm$ 0.02 \\
(13294) Rockox & 2.26 $\pm$ 0.05 & -- & -- & -- \\
(88055) Ghaf & -2.75 $\pm$ 0.15 & -- & -- & -- \\
(23871) Ousha & -2.30 $\pm$ 0.13 & Yes & 3.98 $\pm$ 0.55 & 0.69 $\pm$ 0.02 \\
(59980) Moza & 7.77 $\pm$ 0.21 & -- & -- & -- \\
(269) Justitia & 10.59 $\pm$ 0.10 & -- & -- & -- \\
%Consider using \footnote{\url{http://www.latex-project.org/}} for sources here
\hline
\end{tabular}
\caption{Spectral parameters of the targets produced by the CANA tool \citep{DePra2018}. Slope was measured longward of 0.5 $\mu$m. Targets with a potential 0.7 $\mu$m absorption feature were analyzed for absorption depth between the 0.55 and 0.85 $\mu
$m range.}
\label{tab:CANA}
\end{table*}

\begin{table*}[htb!]
\centering
%\label{tab:tax_table}
\begin{tabular}{lccccccc}
\hline
Asteroid & Bus-DeMeo & SDSS & Tholen 1989 & \added{SkyMAPPER} & \added{PRIMASS} & \added{Mahlke 2022} & This work \\
\hline
(10253) Westerwald & S & S & -- & -- & -- & -- & S \\
(623) Chimaera & -- & -- & X or C & -- & \added{Xc} &\added{ Ch }& X or C \\
(13294) Rockox & S & CX & -- &\added{ X }& -- & -- & X \\
(88055) Ghaf & -- & -- & -- & -- & -- & -- & B \\
(23871) Ousha & -- & C & -- &\added{ C }& -- & -- & B \\
(59980) Moza & X & C or L & -- &\added{ S }& -- & -- & S or L \\
(269) Justitia & D & -- & -- & -- & -- &\added{ Z }& D \\
%Consider using \footnote{\url{http://www.latex-project.org/}} for sources here
\hline
\end{tabular}
\caption{Previous taxonomic classifications for the EMA targets compared to results from this work. SDSS, Tholen 1989\added{, and SkyMAPPER} classifications were made using photometric data. Results from Bus-DeMeo, \added{Mahlke, PRIMASS, }and this work were classified with spectroscopy.}
\label{tab:tax_table}
\end{table*}

\section{Discussion}\label{sec:disc}
We collected and analyzed visible spectra of the EMA mission targets to verify and determine their taxonomic classifications \added{(Table~\ref{tab:tax_table})}. (10253) Westerwald is most consistent with the S-type taxonomy\added{. We infer that Westerwald may be rich in silicates like olivine and or pyoxene due to its absorption feature near 1 $\mu$m \citep{cloutis1986}.} (623) Chimaera and (13294) Rockox fall within the X- or C-type classification\added{, therefore the surfaces of these bodies may be composed of carbonates and/or metals like iron and nickle. Sub-type classification of Chimaera and Rockox cannot be definitively constrained from visible spectra alone and requires NIR observations beyond the wavelength range presented here. NIR observations of Chimaera by \citet{fieber2012} reveal spectral signatures consistent with CM2 carbonaceous chondrites, which are characterized by heavily altered hydrated minerals containing silicates as well as iron and magnesium oxides \citep{rubin2007}. The 0.7 $\mu$m feature present in Chimaera's spectrum is consistent with the presence of oxidized iron and indicates potential aqueous alteration.} (88055) Ghaf and (23871) Ousha show distinct blue-sloped spectra indicative of B-types in the C-complex, \added{meaning Ghaf and Ousha are likely carbonaceous with possible hydrated minerals, volatiles, and organics.}(59980) Moza is likely an L-type asteroid \added{which has been associated with the presence of carbon and organic matter through the spectral comparison with CO and CV chondrites \citep{sunshine2008, mahlke2026}.} (269) Justitia has a distinct red-sloped spectrum indicative of a D-type in the Bus classification. \added{Justitia is likely rich in organics with possible presence of tholins and \added{water} ice. While surface grain size can influence spectral slope and surface property interpretations, we do not account for grain size effects in this work}. All classifications are consistent with the reported albedo of each target, supporting the published diameter estimates and informing mission operations for flyby distance planning and target brightness predictions. 

Justitia's steep red spectrum remains consistent with previous observations \added{\citep{BusBinzel2002,humes2024,mahlke2022}}. This spectral stability is notable given that (203) Pompeja, another main belt asteroid with a similarly steep TNO-like spectrum, has shown measurable spectral variability \citep{hasegawa2022,humes2024pom}. Justitia's consistently steep red slope is characteristic of primitive TNO-like bodies, suggesting a surface with complex organic compounds and \added{potential} origins from the outer solar system \citep{Hasegawa2021}. As the sole rendezvous target of EMA, extended close-approach observations will provide a detailed look at the composition driving this extreme spectral signature. These data will offer new constraints on the origin and evolution of TNO-like asteroids in the main belt.

Our observations show that two of the EMA target asteroids (Ghaf and Ousha) appear to be  B-types. Some of the better known B-type asteroids include (2) Pallas, (24) Themis, (124) Polana, (3200) Phaethon, and (101955) Bennu. Although these asteroids are frequently studied using ground-based observations, Bennu is the only B-type that has been visited and has had surface samples returned via spacecraft. Bennu, the target of NASA’s OSIRIS-REx mission, contains various molecules associated with the building blocks of life, including organic molecules, amino acids, nucleotides, and bio-essential sugars \citep{Glavin2025, furukawa2025}. Bennu was also found to be active and have exogenous boulders \citep{lauretta2019, tatsumi2021}. With similar instruments \citep{edwards2025,filacchione2025,Reed2023}, EMA flyby data of Ousha and Ghaf can be compared to OSIRIS-REx close approach observations of Bennu. Other B-types show distinct characteristics that broaden the context for these comparisons. Water ice has been detected on the surface of Themis \citep{rivkin2010, campins2010}, while Phaethon's exceptionally blue spectrum results from extensive thermal processing that drives surface activity, making it an outlier among B-types \citep{thomas2025,jewitt2010, jewitt2013, li2013}. Future comparisons between VNIR imagery of Phaethon in the JAXA $DESTINY^+$ mission and EMA observations of Ghaf and Ousha will likely highlight these thermal alteration effects.

Close-approach observations of Ghaf and Ousha may reveal both water ice and organic materials, similar to Bennu and Themis. While activity has been observed in B-type NEAs, the mechanisms that trigger such activity may not be present in the main belt \added{\citep{lauretta2019, jewitt2010, jewitt2013, li2013}}. Flyby data from these two targets will enable detailed investigations of compositional similarities and variations within the B-type class. Specifically, EMA observations may address questions about surface activity, exogenous material, and the potential presence of organic compounds. These findings may advance our broader understanding of how primitive asteroids may have contributed to the development of life on Earth. 

Around $\sim$5\% of all asteroids are B-types \citep{demeo2013taxonomic} and have spectra that are most commonly linked to carbonaceous chondrites as their meteorite analogues. \citet{ali2013} found that most B-types display some evidence of a 3~$\mu$m feature, potentially indicating the presence and effects of hydration on many of these bodies. They identified six B-type sub-groups or clusters based on spectral slopes between 0.8 – 2.5 $\mu$m and are proposed to reflect the alteration history based on spectral matches with CI, CM, CV, and CK meteorites \citep{ali2013}. \added{Follow up telescopic o}bservations of Ousha and Ghaf in the VNIR will allow for a more distinctive match within the carbonaceous chondrite group, potentially allowing for interpretation of each asteroid's level of hydration.

The EMA mission will visit 4--6 taxonomic types (S, X, B, D, and potentially C or L), further developing our understanding of the compositional characteristics of a variety of classes. Depending on flyby proximity to each asteroid, EMA may be able to detect potential exogenous boulders on the surface of these targets, informing us about collision rates in the main belt \citep{tatsumi2021}. These flybys will shed light on potential surface heterogeneity and perhaps provide details regarding surface particle sizes. Data from the EMA flybys will provide critical context for interpreting ground-based observations, strengthening the reliability and depth of information that can be inferred from telescopic observations.

\section{Conclusion and Future Work}\label{sec:conclusion}
We obtained visible spectra of all seven EMA targets and find that this group includes C, D, L, S, and X taxonomies. Remarkably, two of these objects are rare B-types, which will provide a unique opportunity to examine more members of this uncommon class in detail. Through close approach observations spanning multiple taxonomic classes, EMA will improve constraints on asteroid composition, collisional evolution, hydration state, surface activity, and spatial homogeneity across diverse asteroid populations.

Future observations of these targets will continue to refine our understanding of asteroid variability and composition. We recommend obtaining NIR and mid-IR spectra of the six flyby targets \added{to confirm and extend the results presented here}. Observations in the 3, 6, and 8--14 $\mu$m range would enable detailed characterization of hydration features and volatile signatures. These data would also allow surface particle size constraints following the approach of \citet{clark2010} and refine taxonomic sub-type classifications, particularly within the B- and X-complexes where 0.8--2.5 $\mu$m observations show more distinguishable differences. Measurements, such as these, might only be possible with JWST. Together, these measurements would provide tighter constraints on the surface composition of each target prior to the EMA encounters (i.e. hydration, volatiles, metals, pyroxene, olivine). Since the EMA instruments will cover these three wavelength ranges, direct comparisons between JWST remote observations and EMA close-approach data can establish ground truth for compositional interpretations of other JWST-observed asteroids. Improved confidence in remote characterization not only extends the scientific reach of both missions, but also strengthens the scientific justification for future mission target selection, particularly for sample return and resource exploration efforts.

Additionally, multi-night lightcurve campaigns for (88055) Ghaf, (23871) Ousha, and (59980) Moza are needed to constrain rotation periods listed in Table~\ref{tab:param_table}. These measurements will inform flyby proximity planning and assess the potential for surface activity.

%% Please use the acknowledgment and contribution environments. This will 
%% be anonomyized when the "anonymous" style option is used. 
\begin{acknowledgments}
We thank the Arizona Technology and Research Initiative Fund at NAU for supporting NAU's access to the Lowell Discovery Telescope, and thank the LDT telescope operators and staff for their support. We respectfully acknowledge that the Lowell Discovery Telescope is located within the Coconino National Forest on the traditional homelands of Indigenous peoples. We also thank the NSF GRFP for supporting this research. \added{We thank David Sand and Bhagya Subrayan for collecting additional photometric observations of (23871) Ousha with the Bok Telescope on Kitt Peak}. We appreciate comments from the EMA science team, including Harish, prior to submission.\\ \added{This work was funded in part by the United Arab Emirates, Emirates Mission to the Asteroid Belt project.}\\
\indent Taxonomic type results presented in this work were determined using a Bus–DeMeo taxonomy classification \added{summary spreadsheet available through the} web tool by Stephen M. Slivan, developed at MIT with the support of National Science Foundation grant 0506716 and NASA grant NAG5-12355.\\
\indent Part of the data utilized in this publication were obtained and made available by the MITHNEOS MIT-Hawaii Near-Earth Object Spectroscopic Survey. The IRTF is operated by the University of Hawaii under contract 80HQTR19D0030 with the National Aeronautics and Space Administration. The MIT component of this work is supported by NASA grant 80NSSC18K0849.\\
\indent This publication makes use of data products from the Near-Earth Object Wide-field Infrared Survey Explorer (NEOWISE), which is a project of the Jet Propulsion Laboratory/California Institute of Technology. NEOWISE is funded by the National Aeronautics and Space Administration.\\
\indent Data from the MPC's database is made freely available to the public. Funding for this data and the MPC's operations comes from a NASA PDCO grant (80NSSC22M0024), administered via a University of Maryland - SAO subaward (106075-Z6415201). The MPC's computing equipment is funded in part by the above award, and in part by funding from the Tamkin Foundation.
\end{acknowledgments}

\facilities{LDT(DeVeny) and Lowell Observatory 0.5m Titan Monitor}

\software{PypeIt, PHOTOMETRYPIPELINE, and CANA 
%\citep{2013A&A...558A..33A,2018AJ....156..123A,2022ApJ...935..167A},  
          %Cloudy \citep{2013RMxAA..49..137F}, 
          %Source Extractor \citep{1996A&AS..117..393B}
          }


\appendix

\section{Additional Tables}

\begin{table*}[htb!]
\centering
%\label{tab:obs_table}
\begin{tabular}{ccc}
\hline
Wavelength ($\mu$m) & Norm. Refl. & Std. Error \\
\hline
0.39 & 0.73 & 0.20 \\
0.40 & 0.62 & 0.28 \\
0.41 & 0.67 & 0.21 \\
0.42 & 0.73 & 0.11 \\
0.43 & 0.81 & 0.11 \\
0.44 & 0.77 & 0.05 \\
0.45 & 0.79 & 0.06 \\
0.46 & 0.82 & 0.03 \\
0.47 & 0.85 & 0.02 \\
0.48 & 0.87 & 0.02 \\
\hline
\end{tabular}
\caption{Mean normalized reflectance spectrum of (10253) Westerwald with corresponding standard error. Entire dataset is available in the online publication.}
\label{tab:WesterwaldData}
\end{table*}

\begin{table*}[htb!]
\centering
%\label{tab:obs_table}
\begin{tabular}{ccc}
\hline
Wavelength ($\mu$m) & Norm. Refl. & Std. Error \\
\hline
0.39 & 0.88 & 0.08 \\
0.40 & 0.91 & 0.03 \\
0.41 & 0.99 & 0.02 \\
0.42 & 0.99 & 0.02 \\
0.43 & 0.95 & 0.03 \\
0.44 & 0.97 & 0.01 \\
0.45 & 0.99 & 0.02 \\
0.46 & 0.97 & 0.01 \\
0.47 & 0.97 & 0.01 \\
0.48 & 0.98 & 0.01 \\
\hline
\end{tabular}
\caption{Mean normalized reflectance spectrum of (623) Chimaera with corresponding standard error. Entire dataset is available in the online publication.}
\label{tab:ChimaeraData}
\end{table*}

\begin{table*}[htb!]
\centering
%\label{tab:obs_table}
\begin{tabular}{ccc}
\hline
Wavelength ($\mu$m) & Norm. Refl. & Std. Error \\
\hline
0.39 & 0.80 & 0.08 \\
0.40 & 0.82 & 0.05 \\
0.41 & 0.84 & 0.04 \\
0.42 & 0.86 & 0.04 \\
0.43 & 0.87 & 0.03 \\
0.44 & 0.90 & 0.03 \\
0.45 & 0.90 & 0.02 \\
0.46 & 0.91 & 0.02 \\
0.47 & 0.92 & 0.02 \\
0.48 & 0.93 & 0.02 \\
\hline
\end{tabular}
\caption{Mean normalized reflectance spectrum of (13294) Rockox with corresponding standard error. Entire dataset is available in the online publication.}
\label{tab:RockoxData}
\end{table*}

\begin{table*}[htb!]
\centering
%\label{tab:obs_table}
\begin{tabular}{ccc}
\hline
Wavelength ($\mu$m) & Norm. Refl. & Std. Error \\
\hline
0.39 & 2.56 & 0.80 \\
0.40 & 2.03 & 0.48 \\
0.41 & 1.78 & 0.37 \\
0.42 & 1.58 & 0.32 \\
0.43 & 1.42 & 0.24 \\
0.44 & 1.16 & 0.09 \\
0.45 & 0.94 & 0.08 \\
0.46 & 0.83 & 0.07 \\
0.47 & 0.93 & 0.06 \\
0.48 & 0.98 & 0.05 \\
\hline
\end{tabular}
\caption{Mean normalized reflectance spectrum of (88055) Ghaf with corresponding standard error. Entire dataset is available in the online publication.}
\label{tab:GhafData}
\end{table*}

\begin{table*}[htb!]
\centering
%\label{tab:obs_table}
\begin{tabular}{ccc}
\hline
Wavelength ($\mu$m) & Norm. Refl. & Std. Error \\
\hline
0.39 & 1.15 & 0.07 \\
0.40 & 1.03 & 0.06 \\
0.41 & 1.02 & 0.04 \\
0.42 & 1.04 & 0.04 \\
0.43 & 1.04 & 0.04 \\
0.44 & 1.04 & 0.02 \\
0.45 & 1.02 & 0.02 \\
0.46 & 1.03 & 0.02 \\
0.47 & 1.03 & 0.02 \\
0.48 & 1.02 & 0.02 \\
\hline
\end{tabular}
\caption{Mean normalized reflectance spectrum of (23871) Ousha with corresponding standard error. Entire dataset is available in the online publication.}
\label{tab:OushaData}
\end{table*}

\begin{table*}[htb!]
\centering
%\label{tab:obs_table}
\begin{tabular}{ccc}
\hline
Wavelength ($\mu$m) & Norm. Refl. & Std. Error \\
\hline
0.39 & 0.49 & 0.12 \\
0.40 & 0.59 & 0.08 \\
0.41 & 0.66 & 0.07 \\
0.42 & 0.70 & 0.06 \\
0.43 & 0.74 & 0.04 \\
0.44 & 0.78 & 0.06 \\
0.45 & 0.81 & 0.05 \\
0.46 & 0.86 & 0.03 \\
0.47 & 0.90 & 0.01 \\
0.48 & 0.93 & 0.03 \\
\hline
\end{tabular}
\caption{Mean normalized reflectance spectrum of (59980) Moza with corresponding standard error. Entire dataset is available in the online publication.}
\label{tab:MozaData}
\end{table*}

\begin{table*}[htb!]
\centering
%\label{tab:obs_table}
\begin{tabular}{ccc}
\hline
Wavelength ($\mu$m) & Norm. Refl. & Std. Error \\
\hline
0.39 & 1.48 & 1.38 \\
0.40 & 1.62 & 1.49 \\
0.41 & 1.71 & 1.51 \\
0.42 & 1.62 & 1.33 \\
0.43 & 1.44 & 1.00 \\
0.44 & 1.42 & 0.90 \\
0.45 & 1.31 & 0.70 \\
0.46 & 1.16 & 0.45 \\
0.47 & 1.06 & 0.27 \\
0.48 & 1.02 & 0.18 \\
\hline
\end{tabular}
\caption{Mean normalized reflectance spectrum of (269) Justitia with corresponding standard error. Entire dataset is available in the online publication.}
\label{tab:JustitiaData}
\end{table*}

%% For this sample we use BibTeX plus aasjournalv7.bst to generate the

\bibliography{sample701}{}
\bibliographystyle{aasjournalv7}



\end{document}